# Development of a patient-specific finite element model of the Transcatheter Aortic Valve Implantation (TAVI) procedure

M. Shirzadi[a,b], B. D. Reddy[b], H. Appa[c]

[a] Department of Mechanical Engineering, University of Cape Town
[b] Centre for Research in Computational and Applied Mechanics (CERECAM), University of Cape Town
[c] Strait Access Technologies, Cape Town, South Africa

Abstract

Transcatheter Aortic Valve Implantation (TAVI) is a procedure developed for replacing the defective aortic valve of a patient as an alternative to open heart Surgical Aortic Valve Replacement (SAVR). The percutaneous nature of TAVI, which is its core advantage in comparison to other SAVR procedures, can however also be one of its main disadvantage. This is due to lack of direct access to the calcified leaflets during the procedure, and hence reliance on the host tissue for the proper positioning and anchorage of the deployed prosthetic valve. This work reports on a study to develop a patient-specific finite element (FE) model of the Transcatheter Aortic Valve Implantation (TAVI) procedure for a patient, using a model of a balloon-expandable percutaneous prosthetic aortic valve as a framework for the prediction of its performance. An experimentally measured left ventricle and aortic pressure profile of a single systolic-diastolic cycle of a resting heart was used for assessing the stent's performance after its deployment. The results of the simulation show that the stent maintains its structural integrity after deployment, and successfully pushes the native leaflets back to keep the aortic root clear of all impediments.

1. Introduction

Aortic Stenosis (AS) is the medical term for the narrowing of the aortic valve due to congenital malformation or degeneration/damage of the leaflet, which is either due to deposition of calcium or acute rheumatic fever that can cause inflammation and thickening of the tissue. The abnormally thick or fused-together leaflets either do not close properly or obstruct the flow of blood, forcing the heart to work harder to compensate [1]. Calcific aortic stenosis, on the other hand, can either occur due to age-related inflammatory degradation of the leaflet tissue, which is being subjected to continuous "wear and tear", or as a result of biochemical and humoral processes [2]. The global prevalence of clinically significant age-related AS in patients over 70 is approximately 1 to 3% and it is estimated that up to 40% of those aged 60 or over, and 75% of those aged over 85 have mild calcified aortic valve [3]. Surgical Aortic Valve Replacement (SAVR) remains the gold standard methodology [4] to treat severe aortic valve disease/defect. However, a new intervention has been developed as a result of a major advance in prosthetic heart valve designs. Transcatheter Aortic Valve Implantation (TAVI) or Transcatheter Aortic Valve Replacement (TAVR) is a new method in which a specially developed, balloon expandable or self-expanding stented prosthetic valve is delivered to the aortic root location, using a catheter, without the requirement of an incision for the exposure of the aortic root. With this revolutionary intervention, the procedure time, complexity, trauma and post-operative rehabilitation period are drastically reduced.

The first human implantation of percutaneous transcatheter aortic valve was performed by Cribier et al. [5]. Since then, several new balloon expandable and self-expanding designs have been developed, such as the Edwards Lifescience SAPIEN and Core valve. All of these valves comprise three leaflets (biological or polymeric) to mimic an actual human tri-leaflet aortic valve. These leaflets are mounted in a stent structure that has the ability to be crimped to a small diameter, in order to be delivered to the deployment site. The balloon expandable valves are usually made of medical-grade stainless steel or cobalt-chromium, and are specifically designed to withstand the large elastic-plastic strains that they are subjected to during the crimping and expansion processes. In addition to this, they are able to tolerate the cyclic biomechanical loading due to cardiac cycle, and have excellent durability against corrosion. The self-expandable valves, on the other hand, are mainly made of Nitinol (Nickel-Titanium), which is a super-elastic metallic alloy with a shape memory property [6].

Currently there are four main approaches by which the prosthetic valve is delivered to the root: transfemorally, transapically, transaortically or via the subclavian artery. Depending on the condition of the patient, surgeons choose the type of valve and delivery approach that has the highest chance of success. Some of the published intraoperative and postoperative adverse events of TAVI procedure include: aortic injury due to excessive radial expansion force during deployment [7], paravalvular leak or impairment of coronary flow or both, stroke, cardiac tamponades [8] and valve migration [9]. From these events, it can be deduced that surgeons are not able to predict reliably the outcome of the procedure due to the nature of the intervention. In the absence of a reliable predictive tool, surgeons are left to rely on personal experiences, guidelines from the prosthetic valve manufacturers and published reports on previous interventions to be able to make decision regarding the optimal valve size and positioning site [10]. It is therefore essential to have a preoperative quantitative understanding of patient-specific biomechanical interaction of the prosthetic valve and the native valve to be able to maximise the chance of success in the TAVI procedure [11]. In recent years, close collaboration between cardiac surgeons and engineers has led to a drive for developing patient-specific computational models of the TAVI procedures that can be used as a reliable predictive tool by surgeons to be able to perform trial deployment simulation and assess the results. Many studies have been conducted aside and in support of this idea in order to create the required frame work. This includes characterization of the relevant tissue behaviour and developing reliable constitutive material models that can be used in the computational models. These may then be used to create patient-specific models using tools such as finite element analysis in order to generate simulations of the procedure. In this regards, the works by Wang et al. [11], Auricchio et al. [12] and Capelli et al. [13] can be mentioned, to name a few. These and many other similar studies have contributed significantly towards achieving an accurate and efficient methodology for computational modelling of the TAVI procedure.

The aim of this research has been to develop an efficient patient-specific finite element simulation of the TAVI procedure using a model of the 23 mm percutaneous prosthetic aortic valve developed by Strait Access Technologies (SAT), for the purpose of its mechanical assessment and post-operative performance. The FE model developed and methodology proposed in this study is intended to serve as a framework for performance assessment of future valve design, and as a step towards realistic simulation of TAVI procedure.

## 2. Methods and materials

*2.1 Patient-specific aortic root model*

The mid-chest area Multi-Slice Computer Tomography (MSCT) images of an 89 years old female patient who was diagnosed with AS was provided by SAT, in the DICOM (Digital Imaging and Communication in Medicine) format, for the purpose of three-dimensional aortic root model extraction. The image processing software ScanIP (Simpleware Ltd, Exeter, UK) was used for this purpose, which allows for 3D image data visualization and model generation [14]. This was done by firstly cropping the aortic root's region and a small section of the left ventricle and ascending aorta from the rest of the image. Several experiments were carried out to determine the optimal grayscale threshold band that allow for the optimal segmentation of the lumen in the aortic root and aid in clutter reduction. The segmented lumen was then extracted as a 3D model and the surrounding unwanted three-dimensional clutter, which was unavoidable during segmentation, were deleted by using a combination of 3D modification tools. The Gaussian recursive filter and several other 3D model processing tools were utilized to smoothen the ragged surface of the 3D model and to further improve its shape. Thereafter, a combination of the available enhancing filters that allow for outlining some of the faint feature in the images were used to improving its visibility. After that the lumen model was duplicated and dilated until it was observed that the aortic root walls were fully covered with little to no overextension. The core lumen model was then deleted. The appearance of the aortic root model was further enhanced and compared with the CT images to ensure that it is as similar as possible to the imaged anatomy. Figure 1 shows the segmented walls of the aortic root form different views along with the extracted model. The short and long axis of the annulus of the patient was measured to be 19.65 and 22.3 mm.

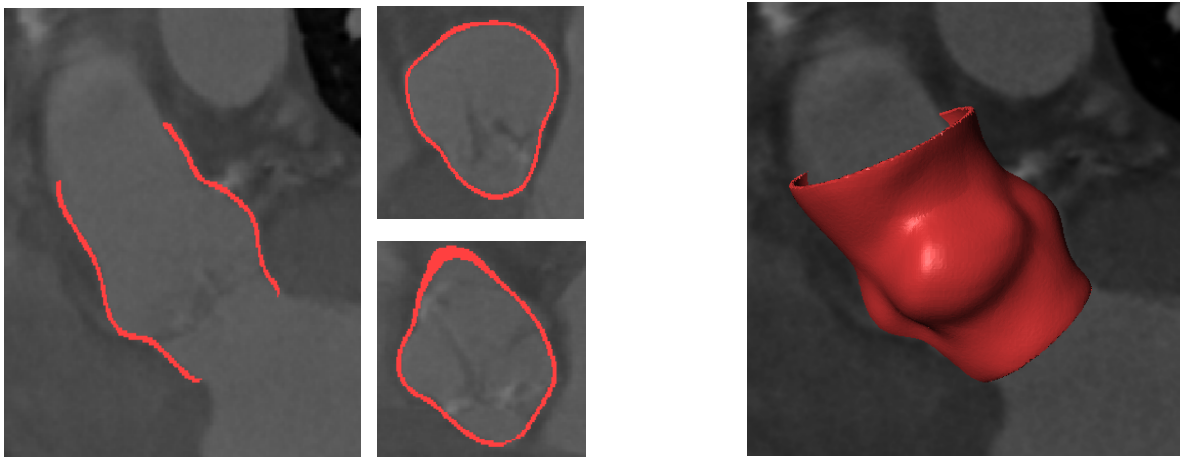

***Figure 1*** *The segmented walls of the aortic root from three different views (left) along with the extracted aortic root model (right).*

This model was firstly converted to a shell model and was then imported into ABAQUS for meshing. The motivation for this conversion was to reduce the cost of the simulations by using shell elements, which provide a good balance between accuracy and efficiency.

It should be mentioned that due to the thin structure of the aortic leaflets, their segmentation and extraction were not possible. Therefore, the native leaflets models were created in SolidWorks (Dassault Systemes, USA), based on the observation from the MSCT images. They were retracted slightly to create an opening configuration to avoid numerical complications due to initial contact with the stent during its recoiling simulation step. Thereafter they were imported into ABAQUS for meshing. Each set of leaflets were then attached to the appropriate root by constraining the degrees of freedom of the nodes on the attaching edges of the leaflet to move with those of the aortic root along this attachment edge. The root and leaflets were meshed with reduced-integrated large-strain quadrilateral shell elements, in order to have an economical simulation and optimally capture the geometry of the models. This is a robust and efficient element and is suitable for a wide variety of applications. However, due to the reduced integration scheme used by this element, the artificial energy of the model, which is added to suppress hourglassing, should be monitored. A mesh sensitivity study that was conducted to choose an optimal average global element size revealed that 0.3 mm allows for both accuracy and efficiency.

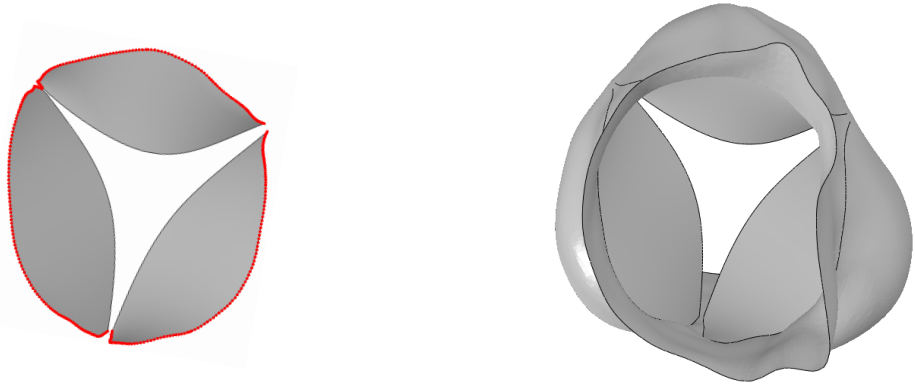

*Figure 2* The view from the left ventricle of the patients' native leaflets, drawn based on the observations from the MSCT images and their attachment to the respective aortic roots. The nodes on the attachment edges that were used to assemble the leaflets on the respective root are shown in red. These leaflets were modelled with the thickness of 0.5 mm.

In order to be able to see the calcific components, the MSCT images were further processed using the available filtering tools in ScanIP. Thereafter, these components were segmented and converted to three-dimensional models in order to assess their proportions, position and individual shapes so that they could be replicated and added to the patient's leaflets models (see Figure 3). No calcific components were observed on the concerned sections of the aortic root walls and were mainly concentrated on the leaflets. These components were meshed with incompatible mode hexahedral element. This element's displacement field is augmented internally with additional displacement modes, which improves it's behaviour when subjected to bending by eliminating the parasitic shear stresses that cause the artificial stiffness in fully integrated elements [15].

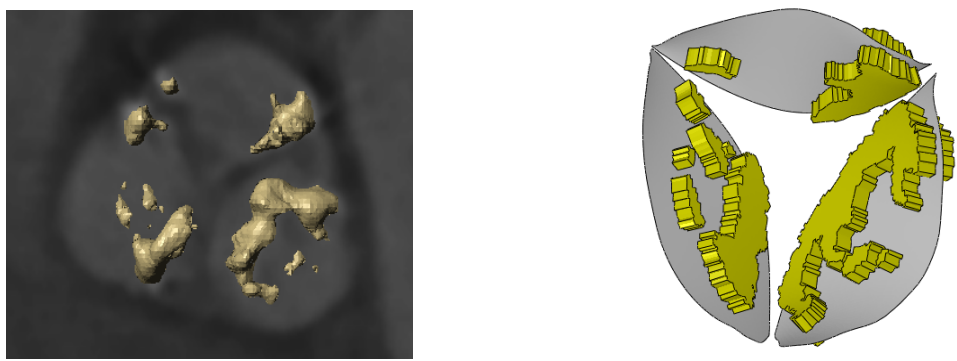

*Figure 3* The identified calcific components that were segmented and converted to 3D model (left) and their FE model on the native leaflets (right)

*2.2 Stent and prosthetic leaflets model*

The 23 mm diameter balloon expandable SAT stent was designed to have one-third cyclic-symmetric geometry. A 3D computer model of its one-third geometry was provided by SAT to be used for its full model generation. The full model of the stent and the prosthetic leaflets were then generated in ABAQUS and meshed. Generally, depending on the geometry of the stent, occasionally they are modelled using beam or shell elements. However, hexahedral elements are more frequently used [11-13, 16, 17] to capture the local peak stress and strain. In this study a similar approach was adapted, and the stent was meshed with the linear reduced integrated hexahedral element. This is due to their ability in providing a good balance between accuracy and computational efficiency. However a major disadvantage of this element is that it is susceptible to hourglassing, which can be mitigated by using at least four elements through the thickness of the structure along with appropriate hourglass controls. The prosthetic leaflets on the other hand were meshed with the linear reduced-integrated large strain triangular shell element, which has no propagating hourglass modes [18] and is better suited for this particular topology and application where the geometry is subjected to large rotations and folding during the crimping and deployment procedure.

*2.3 Material models*

To date, several phenomenological, isotropic and anisotropic, nonlinear elastic material models have been proposed for arterial wall. In this study, the model proposed by Holzapfel et al. (2000) [19], which was further improved by Gasser et al. (2006) [20] by accounting for dispersion of fibres, was implemented for the roots and leaflets, using two and one families of collagen fibres for their tissues respectively [20] (see Figure 4):

$$\overline{\Psi} = c_{10}\left(\overline{I}_1 - 3\right) + \frac{k_1}{2k_2}\sum_{i=1,2}\left\{\exp\left[k_2\left(\kappa\overline{I}_1 + (1-3\kappa)\overline{I}_{4i} - 1\right)^2\right] - 1\right\}. \qquad (1)$$

Here $i$ runs over the number of fibres, $\overline{I}_{4i}$ is the square of stretch in the $i^{th}$ direction, the $c_{10} > 0$ and $k_1 > 0$ are stress-like parameters, and $k_2$ is a dimensionless number. $\kappa$ is a constant that incorporates the effect of fibre dispersion in the model. When $\kappa = 0$, no fibre dispersion is accounted for, i.e. perfect alignment of fibres with the specified directions, and when $\kappa = \frac{1}{3}$ the fibres are randomly distributed, resulting in isotropic response of the tissue. An important assumption in this model is that the fibres are only active in extension and cannot sustain compressive load due to their wavy structure. Hence, the anisotropic part of the strain-energy function would only contribute when the fibres are in extension, i.e. when $\overline{I}_4 > 1$ and $\overline{I}_6 > 1$. If any of these conditions is not satisfied, the relevant part of the anisotropic function is discarded from (1) [20].

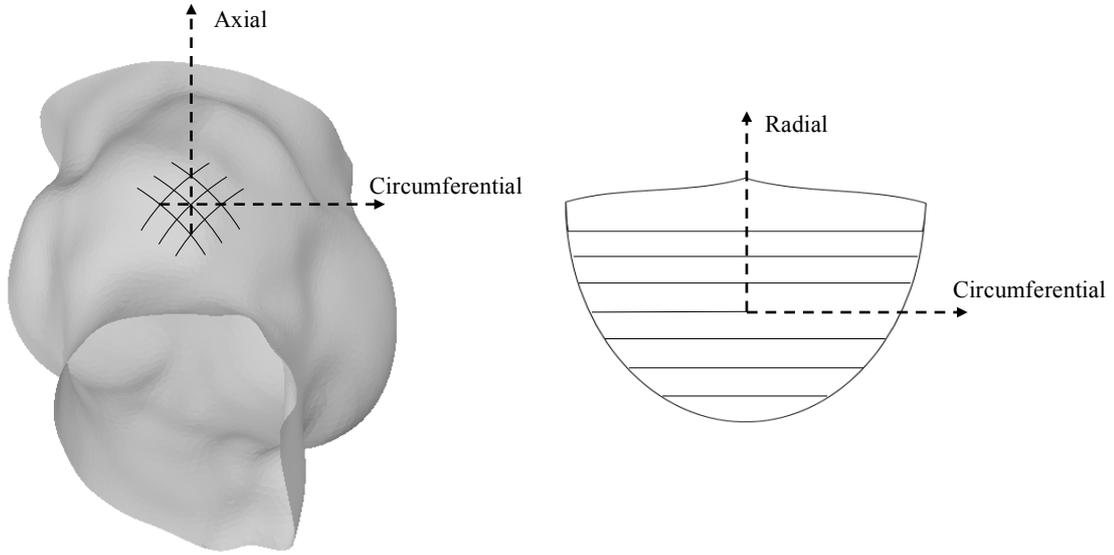

*Figure 4* *The adapted fibre architecture for the aortic roots (left) and native leaflets (right).*

It is assumed that the material and structural properties are constant over their respective geometries and that the fibres are embedded in the plane of the individual elements with no radial component. The calibrated material parameters were acquired from a study by Morganti [21] and are shown in Table 1. In this study, the two families of fibres are assumed to have the same mechanical properties and hence having the same $k_1$ and $k_2$ values.

*Table 1* *The calibrated material parameters for the Gasser et al. (2006) material model acquired from Morganti et al. [21]. Note that $\gamma$ is the defined angle between the two families of fibres.*

| Model | Material Parameters | | | | |
|---|---|---|---|---|---|
| | $c_{10}$ (kPa) | $k_1$ (kPa) | $k_2$ | $\kappa$ | $\gamma$ (deg) |
| Aortic Root | 38.01 | 852.5 | 2245 | 0.22 | 38.5 |
| Leaflets | 41 | 14.71 | 3.83 | 0.05 | - |

The benchmark simulations in [18,20] were replicated and the results were compared to ensure that the implementation of the material models and the fibre directions were correct.

The SAT stent is made of cobalt nickel chromium molybdenum (CoNiCrMo) alloy, originally known as MP35N, which is highly corrosion resistant and has relatively superior fatigue strength that makes it a suitable material for this type of application [22]. In this study, the mechanical response of this alloy is modelled as an isotropic elasto-plastic material, with a linear elastic initial response followed by plastic behaviour with isotropic hardening, using the mechanical properties and stress-strain experimental data shown in Table 2 and 3.

*Table 2* *The mechanical properties of the MP35N alloy used in this study (from SAT).*

| Parameter | Value |
| --- | --- |
| Density | 8415 kg/m$^3$ |
| Young's modulus | 239.5 GPa |
| Poisson's ratio | 0.3775 |
| Yield stress | 350 MPa |

*Table 3* *The post-yield material data of MP35N alloy (from SAT)*

| True Stress(MPa) | True Plastic Strain |
| --- | --- |
| 599 | 0.04879 |
| 704 | 0.09531 |
| 906 | 0.182322 |
| 1090 | 0.262364 |
| 1250 | 0.336472 |
| 1400 | 0.405465 |
| 1520 | 0.470004 |

For the prosthetic leaflets on the other hand, the material evaluation tool in ABAQUS was utilised, which allowed for fitting a suitable material formulation to the uniaxial test data provided by SAT. The isotropic hyperelastic material model proposed by Marlow [23] matched the experimental data optimally and hence was chosen for the leaflets (see Figure 5). This model is given by

$$\Psi = U(J) + \bar{\Psi}(\bar{I}_1)$$

where $\bar{I}_1 = \bar{\lambda}_1^{\,2} + \bar{\lambda}_2^{\,2} + \bar{\lambda}_3^{\,2}$ with $\bar{\lambda}_i = J^{1/3} \lambda_i$ ,

with the $U(J)$ and $\overline{\Psi}(\overline{I}_1)$ denoting the volumetric and deviatoric part of the potential respectively and $\overline{\lambda}_i$ being the deviatoric stretches [18, 23]. In ABAQUS when only one set of data is available for isotropic hyperelastic material, the model proposed by Marlow is recommended to be adapted. In this process, ABAQUS constructs an energy potential to reproduce the uniaxial data exactly, which will perform reasonably well in the other forms of deformations [18]. The version of Marlow's hyperelastic model that is implemented is a special case of the proposed dual strain invariant based potential, where only the first strain invariant is used, allowing for the exact fitting to the experimental data.

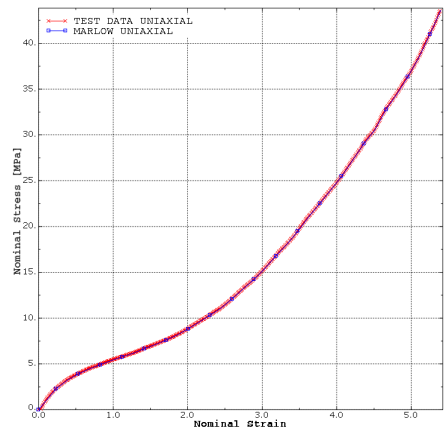

*Figure 5* Curve fitting to the experimental uniaxial test data, using the isotropic hyperelastic material model proposed by Marlow.

## 2.4 Simulation methodology

The ABAQUS/Explicit solver was used for simulating the crimping and expansion of the stent. This was due its ability in handling severely discontinuous forms of nonlinearities, such as complex contact, which makes it an attractive candidate for modelling the stent crimping and expansion processes where very general complicated three-dimensional contact is present in deformable bodies. The SAT stent is crimped to 6 mm diameter before being delivered to the aortic valve location for deployment. The crimper used for this purpose consists of 12 plate-like parts which are designed to apply a radial force to the stent to crimp it to the desired diameter. In this study each of these plates were modelled as a rigid part, which is patterned circumferentially around the stent to a total number of 12 plates. According to the study conducted by Grogan et al. [24] the use of a catheter as a replacement for the balloon to deploy the stent yields similar results for both deformation and stress distribution, whilst eliminating the complexities associated with modelling a balloon and simulating its inflation. Therefore, a catheter was created as a deformable surface for the expansion and deployment of the stent, and was meshed with reduced-integrated quadrilateral surface elements, which is computationally efficient and suitable for this kind of application.

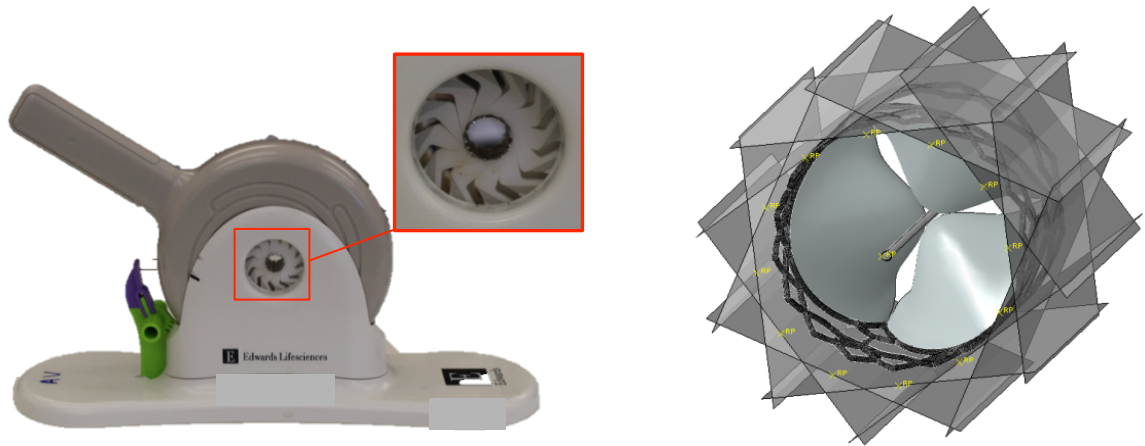

*Figure 6  A photograph of the crimper used in the experiment (left), outlining the configuration of the plates during the crimping process, and the modelled rigid plates patterned circumferentially around the stent to mimic the action of the actual crimper (right).*

In order to be able to assess the performance of the stent after deployment and during its interaction with the aortic root and native leaflets, an experimentally measured left ventricle and aortic pressure profile of a single systolic-diastolic cycle of a resting heart with a rate of 70 bpm was provided by SAT. The pressure profile on the leaflets was acquired by subtracting the ventricular from the aortic pressure profile.

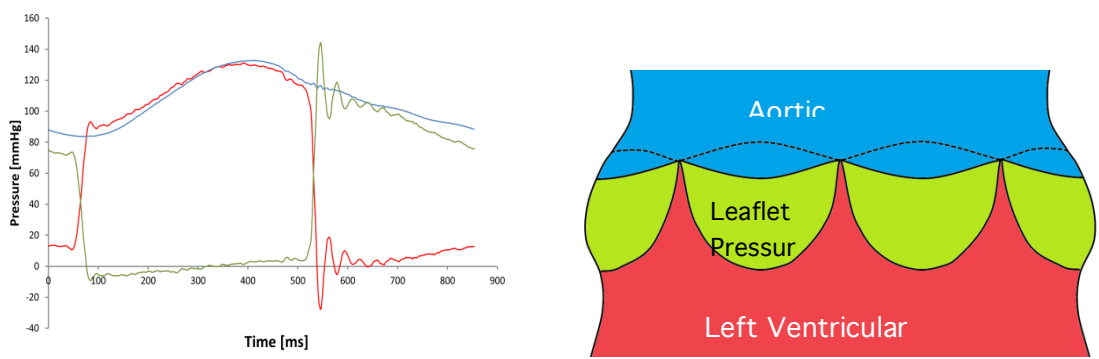

*Figure 7  The pressure profiles applied to the corresponding colour coded aortic region of the heart (left). Note that the leaflet pressure is applied to the prosthetic leaflets during the TAVI simulation.*

After having all the required components, the following simulation steps were created:

(1) crimping of the stent from 23 to 6 mm

(2) removing the crimping plates and allowing the stent to recoil before expansion

(3) expansion/deployment of the stent via the catheter to the designated 23 mm diameter

(4) crimping the catheter to its initial diameter to leave the stent in its expanded state and allow it to go through its second recoiling phase

(5) applying the pressure profiles to their respective surfaces on the aortic root and prosthetic leaflets

3. Results

Figure 8 shows the stress distribution in the deformed configuration of aortic root and native valve at the end of the expansion/deployment simulation step. This is one of the most crucial stages in of the simulations, since at this point the aortic root and leaflet tissues have to stretch to accommodate the catheter and the stent. As can be seen, the stress is mainly concentrated in the interleaflet triangle and annulus regions of the aortic root walls, which are the narrowest sections of the root that has to stretch relatively more in order to expand to the designated deployment diameter of the stent.

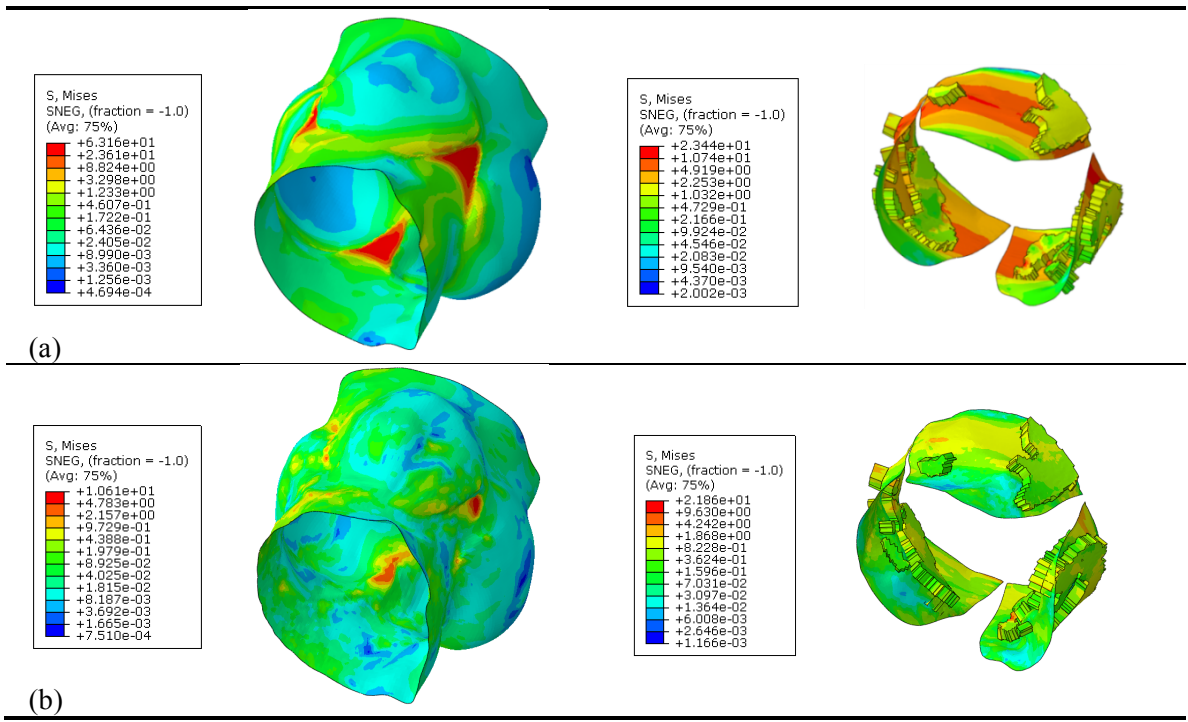

*Figure 8* The stress (logarithmic-MPa) distribution in the deformed configuration of aortic roots and native leaflets at the end of the expansion (a) and relaxation (b) simulation steps.

From the overall position of the stent in the root, it can be seen that no obstruction of coronary ostia is observed and the aortic leaflets are successfully pushed behind. However, some parts of the leaflet belly region along with the calcific components protruded through the openings in the central parts of the stent, but did not impede the opening and coaptation of the prosthetic leaflets (see Figure 9).

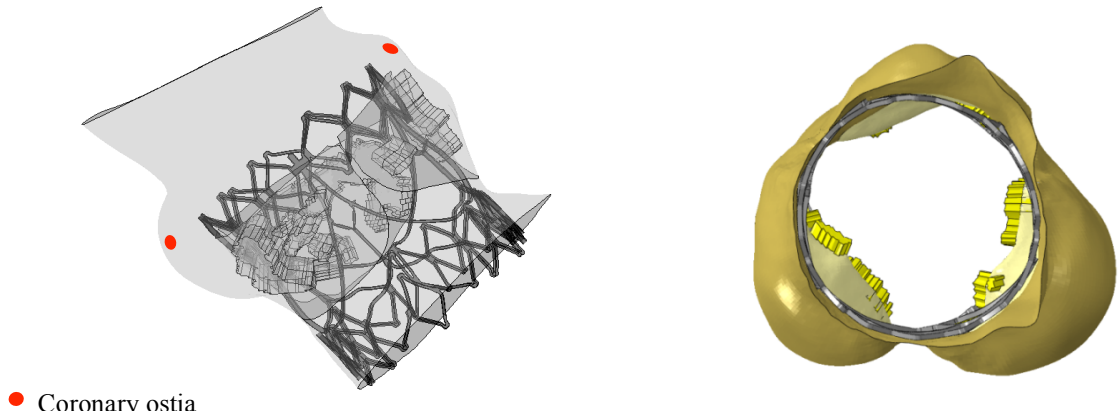

● Coronary ostia

*Figure 9* *The deployed stents in the patient's aortic root after the removal of the catheter and its position with respect to the coronary ostia.*

Figure 10 shows the stress distribution in the prosthetic leaflets during the peak diastolic leaflet pressure in the last simulation step. As can be seen, two major openings were seen to appear at the peak diastolic pressure which might result in paravalvular leakage. During this simulation step, the stent was observed to move towards the left ventricle and back by approximately 3 mm.

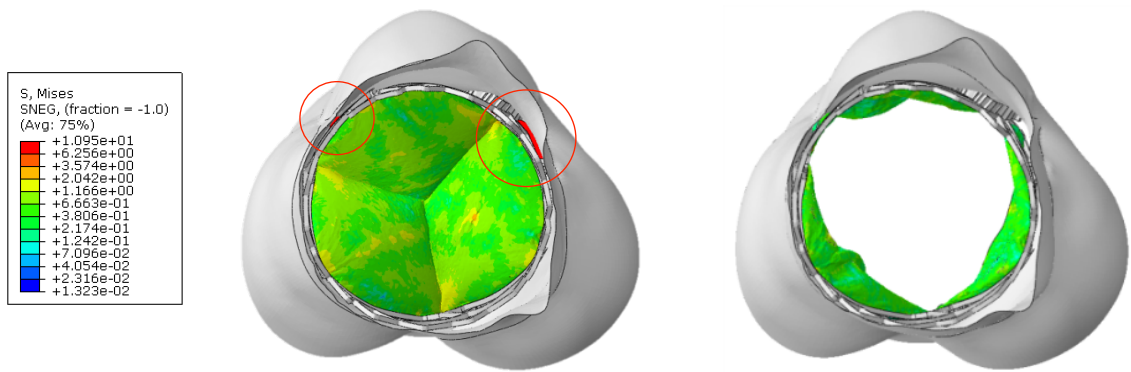

*Figure 10* *The view from the left ventricle of the patient's aortic root, showing the stress (MPa) distribution in the prosthetic valve during the peak diastolic leaflet pressure in the last simulation step, and the possible paravalvular leakage locations (left), and the prosthetic leaflets in their open positions during the systolic phase of the cardiac cycle (right).*

Figure 11 shows the stress distribution in the deformed shapes of the stent at the end of the crimping and expansion simulation steps. The maximum von Mises stress at each individual simulation step is observed to prevail at the complex regions of the stent, i.e. the central and the upper side arms regions. This is due to the fact that these regions have a thinner cross-section and have to endure two phases of large plastic deformations, which makes them more susceptible to fatigue failure. The highest von Mises stress is observed at the end of the expansion simulation step, when the stent has gone through its second plastic deformation process, and is under pressure from both the host tissue and the catheter.

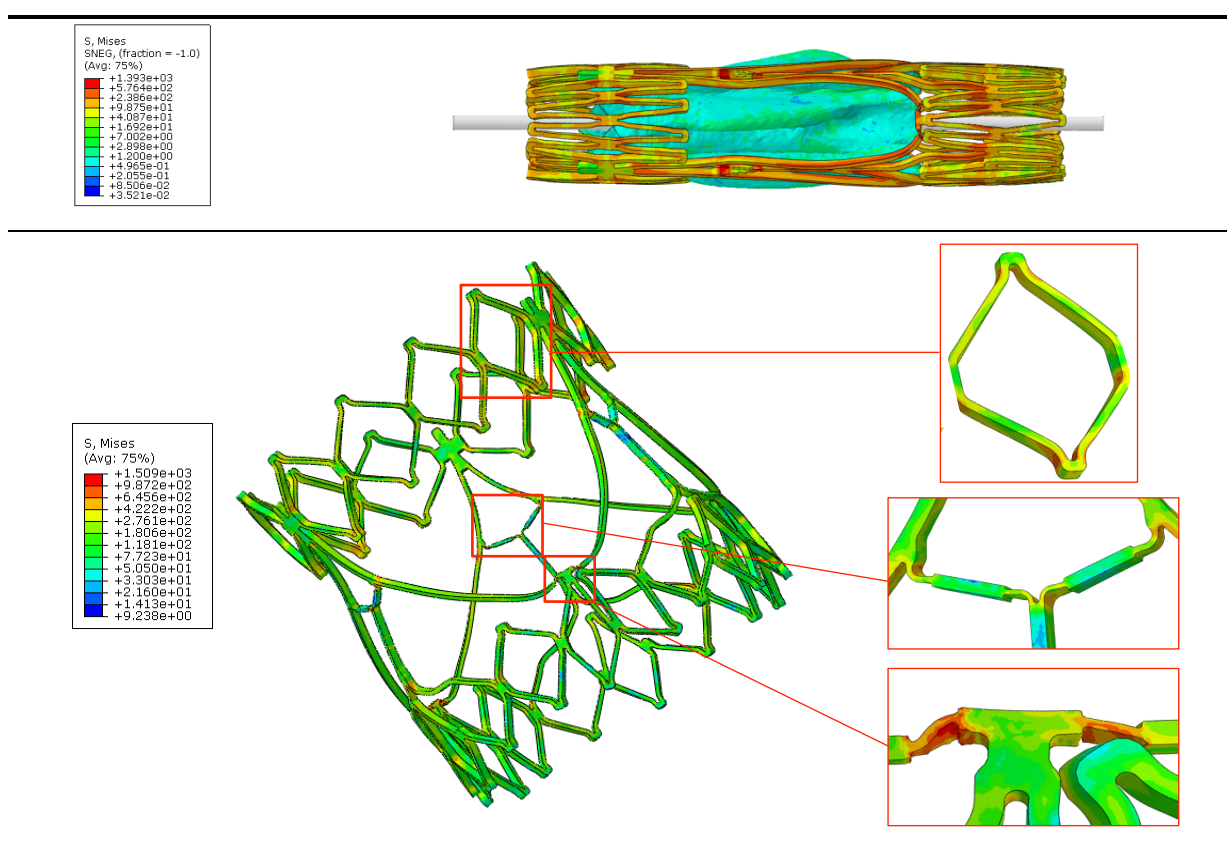

*Figure 11* *The stress distribution (logarithmic-MPa) in the stent geometry after crimping and removal of the crimping plates (a), and after deployment and removal of the cathter, outlining the regions with highest stresses(b).*

## 4. Discussion

The results showed that the stent maintained its structural integrity throughout the simulation, which in turn allowed for the prosthetic leaflets to function normally without any disturbance or impedance even though the topology of the aortic roots and the calcific components had caused it to acquire a somewhat elliptical shape. The prevalence of two major openings outlined in Figure 10 is as a result of the recoiling of the stent and reduction in its diameter, and the expansion of the root under the applied aortic and ventricular pressure, which in turn might result in paravalvular leakage. On the other hand, during the final simulation step, the stent was observed to move towards the left ventricle and back by approximately 3 mm at the peak leaflet pressure during the cardiac cycle. This is mainly due to the elasticity of the tissue and the stent's inadequate anchorage on the native leaflets and aortic root walls, and is an indication that in the long term the stent might migrate and cause further complications. This indicates that the 23 mm valve might not be the correct choice for this patient, and that a larger valve is required to prevent paravalvular leakage and possible valve migration.

The maximum principal stresses in the aortic root and valve tissues were observed to follow approximately the defined collagen fibre directions. This is consistent with the results of previous studies, which implemented an anisotropic material model for the aortic root and valve [25]. In agreement with other studies that have taken a similar approach [21], the stress concentration for the leaflets is observed to be mainly in the belly region and is distributed circumferentially. The circumferentially oriented family of fibres in the leaflet tissue, and the accumulated calcific components that have caused the severe narrowing of the valve regions and act as an obstruction during the deployment process, have led to the stiffer response of the tissue. It was observed that after the removal of the catheter and recoiling of the stent, the stress in the aortic root tissues were substantially reduced.

Although TAVI has obvious advantages in comparison to the classical Surgical Aortic Valve Replacement (SAVR), the lack of reliable predictive tools to assist surgeons in predicting and accounting for unforeseen post-operative adverse events reported in the literature is a concern. The occurrence of most of these events is a direct consequence of improper valve size chosen by surgeons based on experience. Over the past decade, researchers around the world have developed computational models to assist the surgeons in choosing an optimal valve size for each patient and predicting the outcomes of the procedure to increase the chance of success. The model developed in this study is another step towards the improvement of predictive computational models and aid in choosing an optimal diameter valve for the patients. Further, it will serve as a framework for future improvements of the percutaneous aortic valves designs, by allowing for an efficient and accurate simulation methodology.

## 5. Limitations

In this study pre-stress, which is known to exist in arterial tissues, was not included in the developed FE models of the aortic root. Its inclusion would have resulted in a more accurate prediction of the tissue response under the induced expansions. Generally, due to the complexity associated with development of TAVI simulation, inclusion of pre-stress in the aortic root's tissue is almost always omitted, and none of the known similar studies accounted for it. However, for the completeness of the model and increased accuracy of the results, arterial tissue pre-stress should be included in the future. Incorporation of a damage mechanism into the developed FE model was not part of the scope of this project. However, its inclusion will further improve the reliability of the model.

## 6. Conclusion

A patient-specific finite element model of the transcatheter aortic valve implantation procedure was developed for the structural assessment and post-operative performance of the 23 mm percutaneous prosthetic aortic valve developed by Strait Access Technologies (SAT). Structural changes and stress distribution in the aortic root, leaflets and the stent were assessed and the possible risk of valve migration, paravalvular leakage and obstruction of the coronary ostia were evaluated. The simulation methodology proposed and the model developed in this study is a reliable and efficient framework for further improvement of the design of the considered stent. This is another step towards developing a virtual planning tool for the TAVI procedure, which can predict the unforeseen adverse event and increase the chance of success of this new intervention.

## Acknowledgments

This work was funded by Strait Access Technologies (SAT) and was conducted at Centre for Research in Computational and Applied Mechanics (CERECAM) at the University of Cape Town. The authors would like to thank SAT for the support and supplying the required materials.